# Optically induced symmetry breaking due to nonequilibrium steady state formation in charge density wave material 1T-$TiSe_2$

Harshvardhan Jog[1], Luminita Harnagea[2], Dibyata Rout[2], Takashi Taniguchi[3], Kenji Watanabe[4], Eugene J. Mele[5], Ritesh Agarwal[1]*

[1] Department of Materials Science and Engineering, University of Pennsylvania; Philadelphia, PA 19104, USA

[2] Department of Physics, Indian Institute of Science Education and Research, Pune; Maharashtra 411008, India.

[3] International Center for Materials Nanoarchitectonics, National Institute for Materials Science, 1-1 Namiki, Tsukuba 305-0044, Japan.

[4] Research Center for Functional Materials, National Institute for Materials Science, 1-1 Namiki, Tsukuba 305-0044, Japan.

[5] Department of Physics and Astronomy, University of Pennsylvania; Philadelphia, PA 19104, USA

*Corresponding author. Email: riteshag@seas.upenn.edu

**Abstract:**

The strongly correlated charge density wave (CDW) phase of 1T-$TiSe_2$ is being extensively researched to verify the claims of a unique chiral order due to the presence of three equivalent Fermi wavevectors involved in the CDW formation. Characterization of the symmetries is therefore critical to understand the origin of their intriguing properties but can be complicated by the coupling of the electronic and lattice degrees of freedom. Here we use continuous wave laser excitation to probe the symmetries of $TiSe_2$ using the circular photogalvanic effect with very high sensitivity. We observe that the ground state of the CDW phase is achiral. However, laser excitation above a threshold intensity transforms $TiSe_2$ into a chiral phase in a nonequilibrium steady state, which changes the electronic correlations in the stacking direction of the layered material. The inherent sensitivity of the photogalvanic technique provides clear evidence of the different optically driven phases of 1T-$TiSe_2$, as well as emphasizes the interplay of electronic and lattice degrees of freedom in this system under optical excitation. Our work demonstrates that optically induced phase change can occur at extremely low optical intensities in strongly correlated materials, providing a pathway for future studies to engineer new phases using light.

**Significance Statement:** Optical excitation disrupts excitonic correlations in charge density wave system 1T-TiSe2 leading to a structural transition to a chiral structure.

## Main Text:

## Introduction:

Probing the structure and symmetries of a material accurately is fundamental to understanding the origins of their physical properties. (*1–3*) Examining (broken) symmetries often becomes complicated by the interplay between the lattice and electronic degrees of freedom, especially in many strongly correlated materials. (*4–6*) In cases where such subtle symmetry breaking cannot be easily probed using methods typically used to probe crystal structure such as XRD or electron diffraction, the measurement has to be indirect, for example by using selection rules for optical probes to infer the underlying symmetries based on the relevant coupling tensor elements. (*7*, *8*) Charge density wave (CDW) systems are one such class of materials with strong correlations between their electronic and structural degrees of freedom. In CDW systems, the electron density forms periodic modulations at wavevectors $\boldsymbol{q}$ which span the Fermi surface and couple with lattice distortions. (*9–12*) Among these systems, 1T-$TiSe_2$ has been extensively studied, demonstrating an interesting variation of its CDW state, with possible excitonic character leading to claims of excitonic insulating behavior. (*13*, *14*) The material also demonstrates a peculiar triple-$\boldsymbol{q}$ mechanism of CDW modulation, where the CDW has periodic charge modulation along three equivalent wavevectors $\boldsymbol{q}_{1,2,3}$ in the Brillouin zone, which suggest the possibility of chiral ordering in the CDW phase. (*15*) Since the excitonic properties of 1T-$TiSe_2$ are closely related to the structural properties of the CDW phase, it is of great importance to understand the structure of the low temperature phase, which can enable us to gain a better understanding of the origin of its physical properties in particular and other correlated system in general.

1T-$TiSe_2$ is a transition metal dichalcogenide with 2D layers stacked along the $c-$axis where each monolayer is formed by a Ti layer sandwiched between two layers of Se atoms, while the interlayer coupling is via van der Waals forces. The high temperature space group is $P\bar{3}m1$ (Fig. 1(A)). (*16*) 1T-$TiSe_2$ is known to exhibit a $2a_0 \times 2a_0 \times 2c_0$ commensurate CDW transition below ~200K into a structure with $P\bar{3}c1$ symmetry accompanied by a softening of the transverse optical (TO) phonon mode. (*9*, *16–18*) The observed CDW transition is also associated with a possible excitonic condensation, associated with a large binding energy (>50 meV) of a particle hole pair formed by holes in the valence band composed of two Se $4p$ orbitals and electrons in the conduction band with one Ti $3d$ orbital. (*13*, *19*)

1T-$TiSe_2$ has an indirect band overlap of the valence and conduction band above $T_C$, and thus there are three possible Fermi wavevector $\boldsymbol{q_F}$ orientations in the Brillouin zone oriented along the $\overline{\Gamma L}$ directions with an angle of 120° between the three wave-vectors. (*19*) Since $\boldsymbol{q_F}$ determines the direction of modulation of a CDW, with the existence of three equivalent $\boldsymbol{q_F^{1,2,3}}$ wavevectors, there is the possibility of chirality in the CDW phase, where the CDW orientation can wind either clockwise or anticlockwise along the $c-$axis. (*20–22*) However, counterclaims demonstrating the achirality of the CDW ground state in $TiSe_2$ have also been reported. (*23*, *24*) Recent experimental studies have also suggested that femtosecond optical pulses can transiently change the symmetry of 1T-$TiSe_2$, breaking the CDW coherence, from the assigned space group $P\bar{3}c1$ below $T_C$ to $P321$. (*25*, *26*) However, this type of pump probe measurement is not very sensitive to the symmetry and chirality of a material as the response depends on the third order of the permittivity tensor $\chi^{(3)}$, which is non-zero for most systems. However, $\chi^{(2)}$ has stricter symmetry constraints on the allowed elements and is a much more sensitive probe of the symmetry of a material. Thus, in order to probe the symmetry of $TiSe_2$ accurately, an optical probe measuring the $\chi^{(2)}$ response is desirable.

Circular photogalvanic effect (CPGE) and its variants can measure the $\chi^{(2)}$ response of materials and is particularly sensitive to their underlying symmetries and unlike second harmonic generation, the experiments can be performed under very weak continuous wave excitation conditions. (*27–30*) The CPGE technique measures the difference in the DC photoresponse generated by left- and right-circularly polarized light, (*27*) which can only arise in cases where certain crystalline symmetries are broken, making the technique extremely sensitive to spatial symmetries of the material. Under the electric-dipole approximation, CPGE is only allowed in systems with broken inversion symmetry, and more specifically only in gyrotropic point groups, i.e. $C_1, C_2, C_3, C_4, C_6, C_s, C_{2v}, C_{3v}, C_{4v}, C_{6v}, D_2, D_4, D_{2d}, D_3, D_6, S_4, T$ and $O$. (*31*) Based on the symmetries of each of these point groups, the presence of CPGE is determined by the direction of the incoming light and the direction along which the current is measured. Thus, while a crystal having gyrotropic point group symmetry is a necessary condition for observing CPGE, it is not a sufficient condition an depends on the specific experimental geometry, even in the electric-dipole limit.

Here, using CPGE we observe that the ground state of the CDW phase (below $T_C$) is achiral in $TiSe_2$ and belongs to the point group $D_{3d}$, but the system transitions into a chiral phase under continuous wave (CW) illumination above a (relatively low) threshold intensity into the point group $C_3$, indicated by an observed nonzero CPGE. We suggest that this is a result of electronic correlations being suppressed due to an increased free carrier density which screens excitonic correlations in the CDW phase. Our work is also relevant to the question of the excitonic condensate observed in 1T-$TiSe_2$ being disrupted. (*32*) The change in the local symmetry that we report challenges past studies claiming chirality in the ground state of the CDW phase of $TiSe_2$, and are incompatible with earlier theoretical studies where non-local Hall and optical Kerr effects were predicted. (*33*). Importantly, this work also shows that light-matter interactions in correlated phases of matter is highly complex and can be manipulated to a large degree even under very weak optical excitations.

**Results:**

1T-$TiSe_2$ crystals were grown according to previously reported protocols. (*17*, *34*) See Fig. S1 for optical images and Fig. S2 for the X-ray diffraction data of the as-grown crystal. In order to measure the photocurrent response of 1T-$TiSe_2$ of freshly cleaved flakes, we used a modified device fabrication process, which involved prefabricating electrodes and placing the $TiSe_2$ flake on them using PDMS stamping and finally connecting the flake to the prefabricated electrodes with graphene. The entire device was then covered with hBN to protect it from degradation (Fig. 1; see Fig. S3 and Methods section for additional details). The device was cooled down to 140K inside a cryostat for characterization of the CDW state via CPGE measurements.

For a preliminary symmetry characterization of the $TiSe_2$ flake, linear polarized reflectance of the flake was measured at 140K**.** The polarized reflectance data (Fig. 2(A)) shows no change with input light polarization, indicating that the system has crystalline symmetries that do not allow differences between the in-plane ($x - y$ plane) permittivity tensor components. To further test the symmetries of the $TiSe_2$ flake, we measured the polarization dependent photogalvanic response of the device under zero applied bias (see Methods). The incident laser intensity and wavelength were fixed at 100 $\mu$W and 600 nm, respectively, unless otherwise specified. The photogalvanic measurement as a function of changing the angle of the quarter waveplate (QWP) with the incident polarization of the laser (Fig. 2(B)) showed that there is a difference in the response at 45° and 135°, which correspond to left-circularly polarized (LCP) and right-circularly polarized (RCP) light, respectively. This indicates the presence of CPGE in the system. The data can be fit to the phenomenological equation for CPGE given by, (*27*)

$$V_{tot} = V_{CPGE} \sin(2\theta) + V_{LPGE} \sin(4\theta + \delta) + V_0 \quad (1)$$

where $V_{tot}$ gives the total photovoltage (Fig. 2(B)), $V_{CPGE}$ and $V_{LPGE}$ give the circular and linear photogalvanic effect coefficients respectively, $\delta$ is a phase factor, while $V_0$ gives the background photocurrent, usually of photothermal and other origins.

However, in order to measure the PGE response with more speed and accuracy, we used a modified fast-rotating QWP setup (see Methods). (*35*) The laser was maintained at normal incidence to the sample, equidistant from the graphene electrodes to avoid any Schottky effect related response or spatially dispersive CPGE. (*29*, *36*) Note that under normal incidence with photovoltage measurement in the $x - y$ plane, only the point groups with no in-plane $C_2$ symmetry, i.e., $C_1, C_{1v}, C_3, C_{3v}$, allow nonzero off-diagonal elements in the gyrotropy tensor and can therefore produce CPGE, since they permit coupling of purely in-plane electric fields to in-plane currents. However, while $TiSe_2$ belongs to the point group $D_{3d}$, which is inversion symmetric, surprisingly the photoresponse measurement showed a non-zero signal at the $2\omega$ frequency, indicating presence of CPGE at normal incidence (Figs 2b). Significantly, we note that no "training" of the $TiSe_2$, i.e., cooling the device while being illuminated with either LCP or RCP light as mentioned in a previous report, (*22*) was required for the appearance of CPGE.

To identify the origin of the CPGE response of the $TiSe_2$ device, we performed temperature dependent measurements (Fig. 2(C)), which shows non-zero CPGE at low temperatures but drops sharply at 220K to eventually become zero as the temperature increases. This implies that the CPGE signal is an intrinsic property of the CDW phase of $TiSe_2$ and disappears in the high temperature phase where no CDW exists. Note that the CDW transition temperature is ~$190-200$K in 1T-$TiSe_2$ ($T_C = 195$K for the crystal used here, see fig. S4), but hBN encapsulation can increase this transition temperature up to $\sim 230$K. (*37*) Furthermore, the small dip in the CPGE signal at $\sim 175$K can be explained by noting that the resistivity of 1T-$TiSe_2$ (Fig. S4) increases to reach a maximum at $\sim 175$K, and then drops as the temperature increases, due to a crossover from hole-dominated to electron-dominated conductivity. (*38*)

To further probe the system, we performed wavelength dependent CPGE measurements normalized with respect to the input laser intensity (Fig. 2(D)). The CPGE spectrum shows a sharp peak at 450 nm with two smaller peaks at ~550 nm and 640 nm. Importantly, the Lorentzian lineshape of the wavelength dependence indicates a dipolar origin of the CPGE, as opposed to a quadrupolar CPGE (QCPGE) mechanism, which would instead have an anti-resonant lineshape. (*30*) However, while a dipolar origin can be assigned to the observed CPGE, it is incompatible with the assigned point group symmetry of $TiSe_2$ ($D_{3d}$) which forbids CPGE in the dipole approximation due to the presence of inversion symmetry.

In order to resolve the apparent contradiction of the anomalous CPGE observed under CW excitation in our experiments, which is typically of much lower intensity than pulsed excitation, we measured the laser intensity dependence of the CPGE signal at 600 nm wavelength (Fig. 3(A)). As seen in Fig. 3(B), we observed three distinct regimes of the CPGE response as a function of the laser input intensity. The low intensity region ($I \leq 10\mu$W for device shown in Fig. 3(B)) shows a non-zero DC photovoltage but no measurable CPGE. At a threshold intensity $I_T \sim 10\mu$W, there is a sudden onset of the CPGE response, with the signal rising superlinearly with increasing laser intensity. In the third regime above $\sim 30\mu$W, the CPGE showed a linear dependence on the input intensity with a reduced slope. No hysteresis of the signal was observed when the laser was cycled between high and low intensities. Similar behavior was seen in multiple devices observed with different values of $I_T$ depending on slight differences in the device geometry (see Fig. S6 for data from a different device).

The linear photogalvanic effect (LPGE) response, i.e, linear polarization-dependent photoresponse, also requires broken inversion symmetry but has fewer symmetry constraints than CPGE, (*39*) and shows similar behavior as CPGE with laser intensity (Fig. 3(C))**,** with a non-zero value in LPGE only appearing for $I > I_T$. Thus, while LPGE would not be allowed in $D_{3d}$ similar to CPGE, the presence of LPGE response at higher intensities indicates that similar light-induced symmetry breaking processes are governing the presence of CPGE and LPGE in $TiSe_2$.

**Analysis:**

We summarize our results of the CDW phase in 1T-$TiSe_2$ into four main observations : 1) isotropic reflectance in the linear response at low temperatures, 2) presence of PGE in the low temperature phase but absence of PGE in the high temperature phase, 3) Lorentzian lineshape of the wavelength dependence of observed CPGE, and 4) non-linear intensity dependence of the observed CPGE (and LPGE) with threshold behavior. The isotropic nature of the input polarization-dependent reflectance spectrum indicates that the combined $TiSe_2$-Graphene-hBN system does not have any unique crystallographic axis in the $x-y$ plane, which can potentially arise due to strain or Schottky fields due to electrodes or other factors in the material under study. We can therefore eliminate the point groups $C_1, C_i, C_{1v}, C_2,$ or $C_{2v}$ which would manifest as an anisotropic pattern in the reflectance spectrum due to the presence of two unequal axes in the $x-y$ plane. Hence, consistent with the observation of zero CPGE at low intensities, the CDW phase of $TiSe_2$ is assigned a symmetry $D_{3d}$, since the low intensity white light does not coherently excite carriers to cause disruption to the ground state CDW. Also, even under low laser intensity ($I < I_T$), we assign the group symmetry to be $D_{3d}$.

The presence of CPGE under normal incidence of laser at high intensities on the CDW phase of $TiSe_2$ however restricts the possible symmetry of the underlying system to point groups with off-diagonal elements in the gyrotropy tensor $\gamma$ (supplementary section 1). Thus, $D_{3d}$ cannot be the symmetry of the system under higher intensity excitation ($I > I_T$) since the inversion symmetry in $D_{3d}$ prohibits the presence of dipolar CPGE in the system. Moreover, the proposed optically induced point group $D_3$ as discussed in ref. (*26*) does not explain the observed CPGE response either, since the $C_2$ symmetry with the rotation axis in the $x-y$ plane forbids any coupling of the incident electric fields $E_{x,y}$ with the photoresponse $V_{x,y}^{CPGE}$ being measured in the current device geometry (see supplementary section 1).

The most important observations are the nonlinearity of observed signal with laser intensity and the observation of the non-zero threshold intensity $I_T$ for the signal (Figs. 3(A)-(B)). The system showed no hysteresis upon cycling through high and low intensity laser excitation, indicating that there is a sudden optically induced phase transition in the system at higher-intensity laser excitation, likely of electronic origin. For $I < I_T$, the CDW phase of $TiSe_2$ is in its ground state with the point group $D_{3d}$ which does not support any CPGE or LPGE. However, when the laser intensity is increased to $I > I_T$, this is no longer true. Optical excitation of $TiSe_2$ above a critical fluence threshold leads to generation of a free carrier density in the conduction band. This affects the coherence between CDWs on different layers along the $c-$axis due to a transient breakdown of the excitonic correlations formed between the Ti $3d$ orbital and the Se $4p$ orbitals. (*25*, *32*) DFT calculations (*26*) have shown that the ground state CDW of $TiSe_2$ can transiently transform into the space group $P321$ with femtosecond excitations. The space group $P321$ corresponds to the point group $D_3$ where the $C_2$ symmetry with rotation axis in the $x-y$ plane is preserved for a monolayer, setting the $+z$ and $-z$ directions to be equivalent; however, this is not the case for realistic $TiSe_2$ crystals with multiple layers. Due to the Thomas-Fermi screening from the optically generated free carriers, the CDW is suppressed, and there is no phase correlation between the CDW of any particular layer to that on the top and bottom layers. (*26*) Thus, the charge densities at a distance $+z$ are not equivalent to that at $-z$ from any given point of origin, thereby breaking the $C_2$ symmetry with its rotation axis in the $x-y$ plane. Since the electron densities at $+z$ and $-z$ are now distinct due to the broken symmetry, a net local polarization is formed, and the overall symmetry of the system reduces to $C_3$. Thus, the point group of the bulk should be $C_3$ instead of $D_{3d}$ in the presence of optical illumination, when $I > I_T$. The point group $C_3$ belongs to the polar point groups and has non-zero off-diagonal elements in the gyrotropic tensor necessary to produce a dipole-order CPGE by coupling the incident in-plane electric fields to the generated photoresponse (see supplementary section 1).

The nonlinear behavior of the observed CPGE with laser intensity as well as the requirement of a threshold intensity $I_T$ for a nonzero CPGE response suggests that the change in the local phase to the symmetry broken $C_3$ phase tracks the free carrier generation and subsequent breakdown of the CDW order along the $c-$axis (or the $z-$axis in the lab frame). The observed trend is reminiscent of a three-level system under optical pumping, leading to a free carrier population forming in an excited state under non-equilibrium. (*40*) When the laser illumination is turned off, the optically excited free carriers quickly relax, and the system returns to the inversion symmetric ground state. The process has been schematically summarized in Fig. 3(D), showing the coherent CDW ground state, followed by disruption in the CDW coherence due to optically excited carriers, and finally turning off of the illumination resulting in a return to the ground state. We can thus model the observed intensity dependence of the CPGE (and LPGE) in Figs. 3(A-C) as a driven three-level system (see supplementary section 2):

$$V_{CPGE} = \frac{(1-a)I - 1}{(1+2a)I + 1} b + cI + d \tag{2}$$

where $I$ is the intensity of light, and $a, b, c, d$ are fitting parameters. The first term denotes the superlinear increase of CPGE in the low intensity region of the plot, which is similar to what is observed in a three-level system where optical excitation can lead to the formation of a non-equilibrium steady state (NESS) formation with a quasi-stable free carrier population in the conduction band (CB). (*32*) The fitting parameter $a$ is related to the ratio of the relaxation times between the higher energy states to CB and from CB to valence band (VB), and the fit in Fig. 3(A) gives a value of $a = 6.6 \times 10^{-8}$ which indicates a fast decay from the excited state to the bottom of the CB. Such a fast decay is necessary to generate a non-equilibrium population of free carriers for a transient breakdown of the CDW phase. At higher intensities, the equation reduces to $V_{CPGE} \approx cI + d$, implying that after the non-equilibrium steady state is created in the conduction band bottom, further excitation of the system will only increase the CPGE response without further increasing the non-equilibrium free carrier population in the CB. Interestingly, the LPGE is also activated at the same threshold intensity as CPGE (Fig. 3(C)), which implies that the ground state of the CDW before the threshold is symmetric, since the symmetry requirements for non-zero LPGE are more relaxed as compared to those for CPGE. (*39*) The point group $D_{3d}$ satisfies this requirement, thus confirming that the ground state of the CDW state in $TiSe_2$ is achiral, while $C_3$ being a polar group allows both CPGE and LPGE, as we observe above a threshold laser intensity.

## Conclusions:

Carefully designed temperature, wavelength, and intensity dependent CPGE measurements that are very sensitive to the symmetries of the material, show that the ground state of the CDW phase in $TiSe_2$ is achiral due to the absence of any CPGE or LPGE at low intensity optical excitation. However, exciting $TiSe_2$ above a threshold laser intensity leads to an optically driven phase transition into a chiral structure, triggering CPGE and LPGE response. The effect of light on crystalline symmetry is strong in $TiSe_2$ due to the importance of electron-electron and electron-phonon correlations to the stability of the ground state of the system, and thus any disruption of these correlations due to free carrier generation leads to a breakdown in the symmetries of the material. Our work thus also sheds light on the role that excitonic correlations play on the symmetries of $TiSe_2$. An important point also to be highlighted is that light can probe and disrupt electronic correlations even at very low intensities and can be utilized to create novel phases of matter.

**Acknowledgements:**

This work was supported by the US Air Force Office of Scientific Research (award# FA9550-20-1-0345). This work was partially supported by the King Abdullah University of Science & Technology (OSR-2020-CRG9-4374.3) and NSF through the University of Pennsylvania Materials Research Science and Engineering Center (MRSEC) (DMR-1720530) seed grant. Device fabrication and characterization work was carried out in part at the Singh Center for Nanotechnology, which is supported by the NSF National Nanotechnology Coordinated Infrastructure Program under grant NNCI-1542153.This work was supported by L.H. acknowledges financial support from DST-India (DST/WOS-A/PM-83/2021 (G)) and IISER Pune for providing the facilities for crystal growth/characterization. D.R. is thankful for support from IISER Pune. K.W. and T.T. acknowledge support from the JSPS KAKENHI (Grant Numbers 19H05790, 20H00354 and 21H05233).

**Author Contributions:**

HJ and RA conceptualized the project. LH and DR synthesized single crystal 1T-$TiSe_2$ and performed XRD and electrical resistivity characterization. TT and KW synthesized hBN. HJ, EJM, RA formulated the experiments and performed theoretical investigation. HJ fabricated the devices and performed all optoelectronic measurements. EJM, RA acquired funding for the project. HJ, RA drafted the paper, and all authors contributed to reviewing and editing the final draft. RA supervised the project.

**Competing interests:** Authors declare that they have no competing interests.

**Data and materials availability:** All data are available in the main text or the supplementary information.

**Figures:**

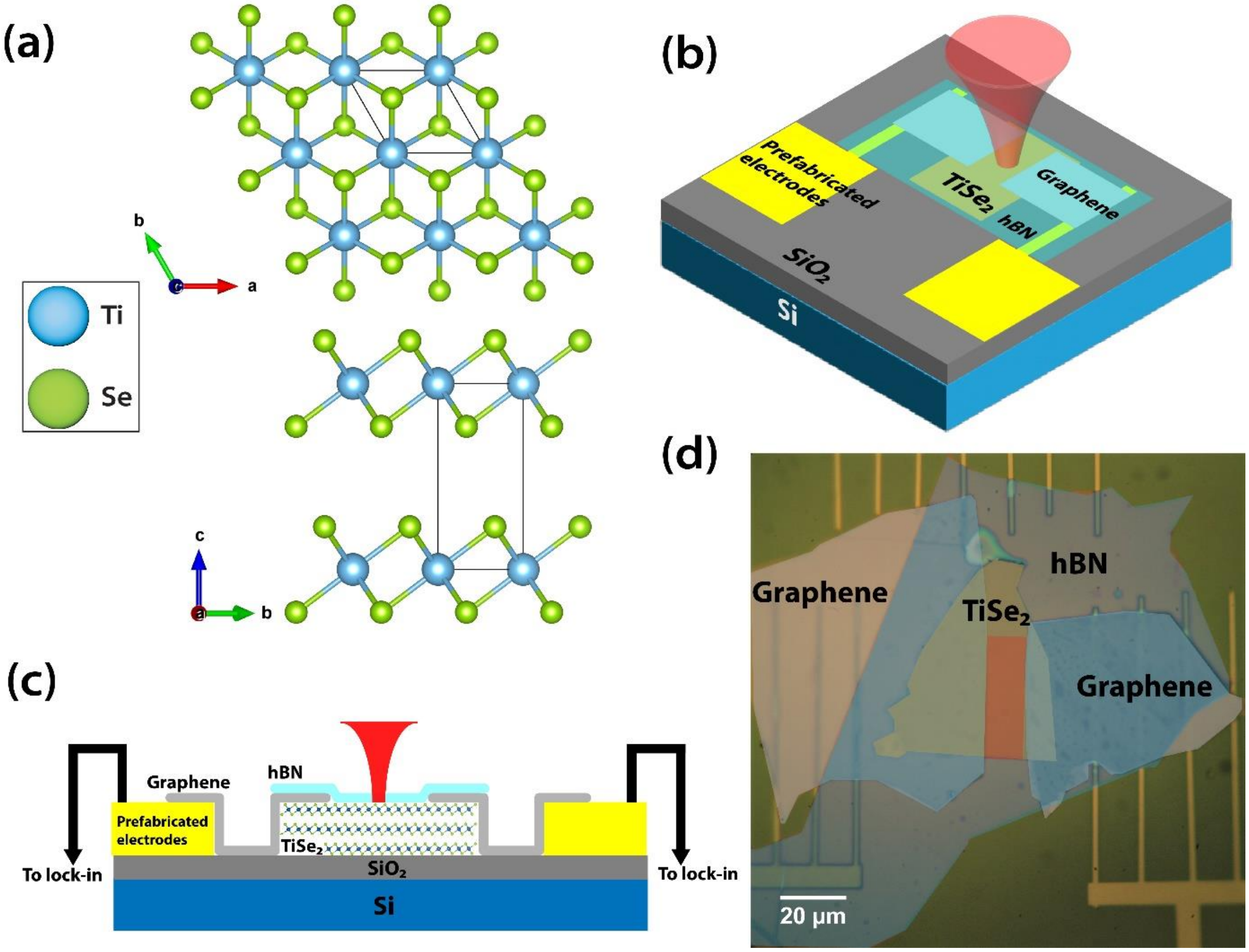

**Figure 1. Crystal structure and TiSe$_2$ device architecture used for photocurrent measurements:** (a) Crystal structure of 1T-TiSe$_2$ showing the view along the $c-$ (top) and $a-$axis (bottom). The material has Van der Waals bonding along the $c-$axis, making it quasi-2D in nature. (b) Schematic of a device as fabricated for this study. See the Methods section for more details. (c) Cross-sectional schematic of the device architecture. (d) A false color image of a typical device used. The electrodes are multi-pronged to ensure good contact with graphene flakes. The red shaded region in the center marks the active device area.

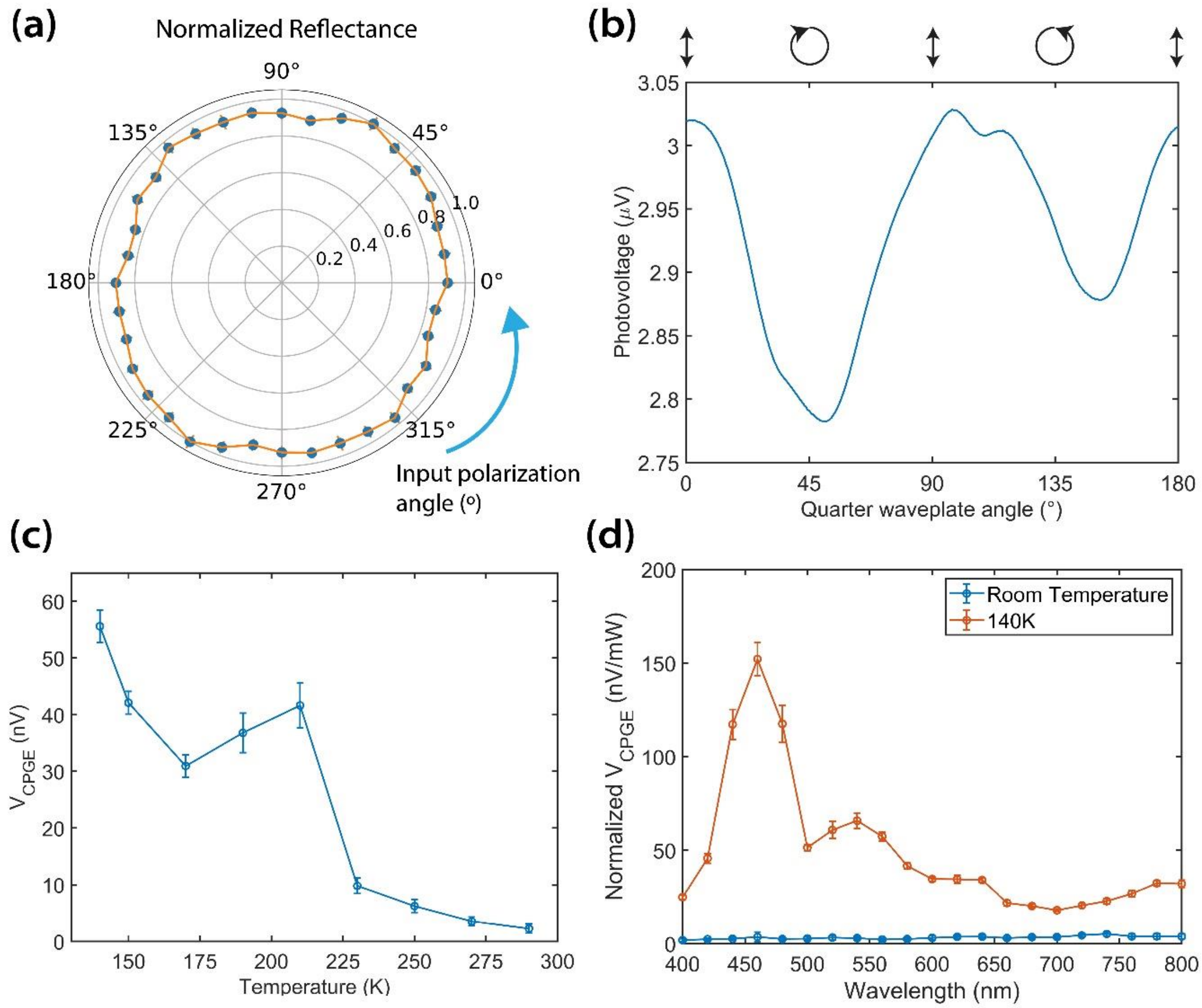

**Figure 2. Device characterization using linear and nonlinear photocurrent measurements to probe symmetry of the CDW phase in $TiSe_2$:** (a) Normalized reflectance of $TiSe_2$ from the active device area showing no unique axis in the $x - y$ plane. (b) Typical photoresponse curve of a $TiSe_2$ device as a function of quarter wave-plate angle with the input laser polarization. The device responds differently to left- and right-circularly polarized light, signifying the presence of CPGE. (c) Temperature dependence of $V_{CPGE}$ in a $TiSe_2$ device. The CPGE is non-zero in the low temperature region while decays sharply above the transition temperature. The dip at $\sim$175K is due to an increased resistivity of $TiSe_2$ at that temperature (see Fig. S4). (d) Wavelength dependence of $V_{CPGE}$ normalized with input intensity, showing sharp Lorentzian peak at $\sim$450 nm and smaller peaks at $\sim$550 nm, 650 nm. The Room temperature CPGE curve is uniformly zero.

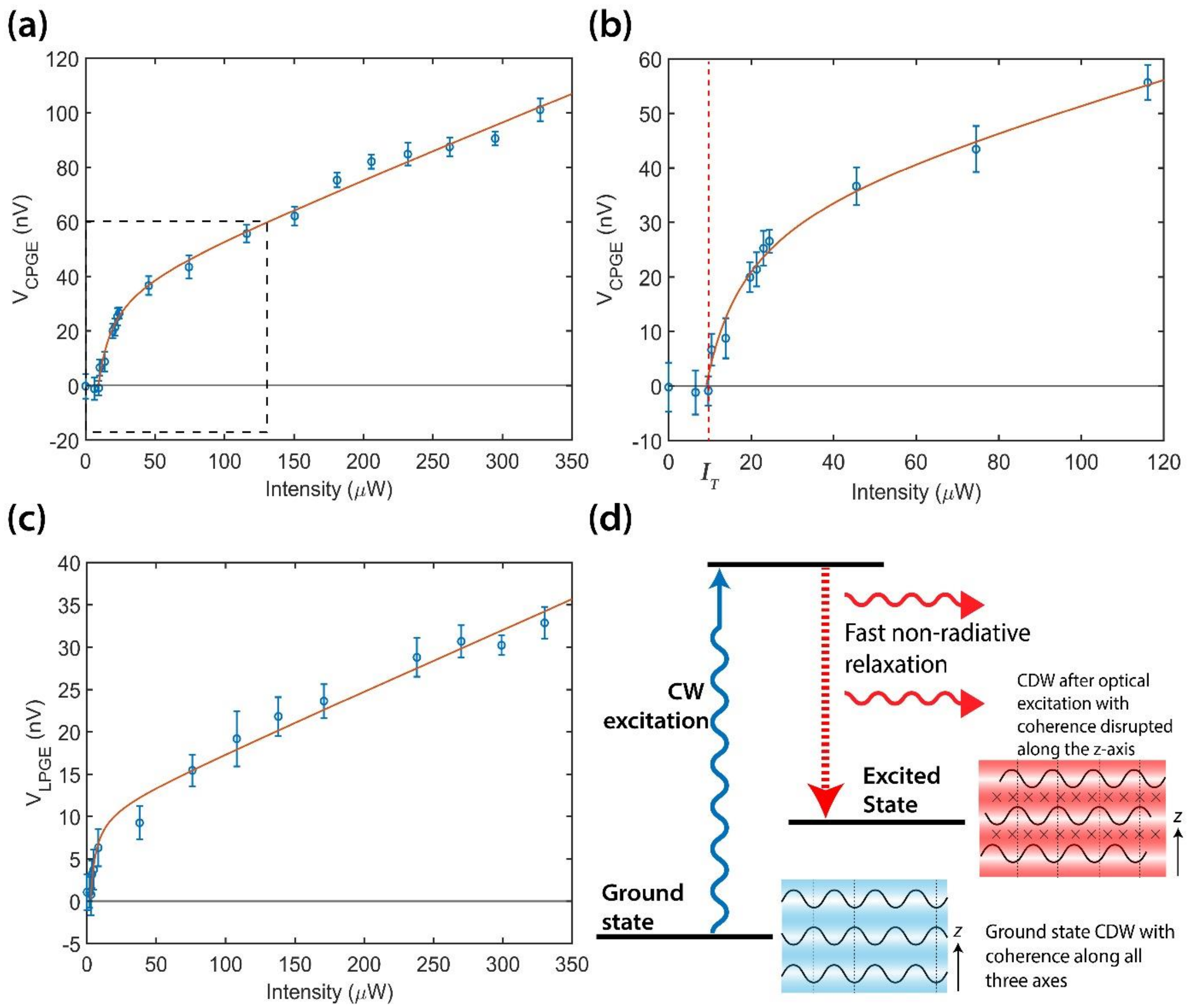

**Figure 3. Intensity dependence of the PGE showing nonequilibrium behavior in $TiSe_2$ below $T_C$.** (a) Intensity dependent measurement of CPGE for a typical $TiSe_2$ device. The red line denotes a fit to the non-linear intensity dependent equation (2) in the main text. (b) Low-intensity region of (a) [the region in the dashed box] zoomed in to show the optically activated nature of CPGE in the system. The observed CPGE is zero up to the threshold intensity $I_T \sim 10\mu$W, then sharply increases, and eventually the slope reduces to a stable value. (c) Intensity dependence of LPGE, showing similar optically activated behavior above the threshold intensity. (d) Schematic of possible process explaining optically triggered behavior. Turning laser illumination on creates a nonequilibrium steady state with free carriers in the conduction band, disrupting the coherence of the CDW in the $z-$ direction, reducing the overall symmetry to $C_3$.

# Supplementary Materials for

Optically induced symmetry breaking due to nonequilibrium steady state formation in charge density wave material $1T\text{-}TiSe_2$

Harshvardhan Jog[1], Luminita Harnagea[2], Dibyata Rout[2], Takashi Taniguchi[3], Kenji Watanabe[4], Eugene J. Mele[5], Ritesh Agarwal[1]*

[1] Department of Materials Science and Engineering, University of Pennsylvania; Philadelphia, PA 19104, USA

[2] Department of Physics, Indian Institute of Science Education and Research, Pune; Maharashtra 411008, India.

[3] International Center for Materials Nanoarchitectonics, National Institute for Materials Science, 1-1 Namiki, Tsukuba 305-0044, Japan.

[4] Research Center for Functional Materials, National Institute for Materials Science, 1-1 Namiki, Tsukuba 305-0044, Japan.

[5] Department of Physics and Astronomy, University of Pennsylvania; Philadelphia, PA 19104, USA

*Corresponding author. Email: riteshag@seas.upenn.edu

**This PDF includes:**

## Methods

### 1. Materials growth and characterization

Single crystals of $TiSe_2$ were grown using iodine vapor transport with Se excess as described in refs (*17*, *21*) of the main text. Ti (99.9 %, metals basis, Alfa Aesar) and Se (99.999+%, metals basis, Alfa Aesar) powders, taken in a molar ratio of 1: 2.2, along with small amounts of iodine (5 mg/cm$^3$), were loaded into a quartz ampoule (≈ 40 cm$^3$). The ampoule was then evacuated, sealed, and placed in a three-zone horizontal oven, where a temperature gradient of 100 ºC was maintained between the source and crystallization zones, for 2 weeks. The growth temperature was set to 600 ºC, to minimize deviations from stoichiometry, respectively, Ti incorporation which occurs as the growth temperature increases [see ref. (*17*) of the main text]. The as-grown crystals are shaped as thin plates, with lateral dimensions of a few mm and thicknesses under 0.5 mm. The single crystals were characterized using a series of techniques such as X-ray powder diffraction (Bruker D8 diffractometer (Cu $K_\alpha$ radiation); X-ray single crystal (Bruker, KAPPA APEX II CCD DUO, Mo $K_\alpha$ radiation), HRTEM (JEOL JEM 2200FS 200 keV), scanning electron microscope (ZEISS GeminiSEM 500) equipped with an energy dispersive X-ray spectroscopy probe (EDX), resistivity measurements, etc. The single crystals are single phase, crystallizing in the space group P-3m1 (164) (JCPDS – 00 – 030 - 1383), with lattice parameters at room temperature of a = 3.5472 Å and c = 6.0183 Å, corresponding to a 1T–$TiSe_2$ structure. The experimental observations indicate a high single crystal quality. The chemical composition was determined by recording EDX data over several areas for a few single crystal pieces. Furthermore, the chemical mapping was performed, and the samples were observed in backscattered-electron mode (BSE) to estimate their homogeneity. The samples proved to be homogeneous and stoichiometric within the error bar of the technique (1 at. %) with a Ti : Se ratio of 0.99 ± 0.01: 2.01 ± 0.01, where 0.01 represents the standard deviation as determined from measurements.

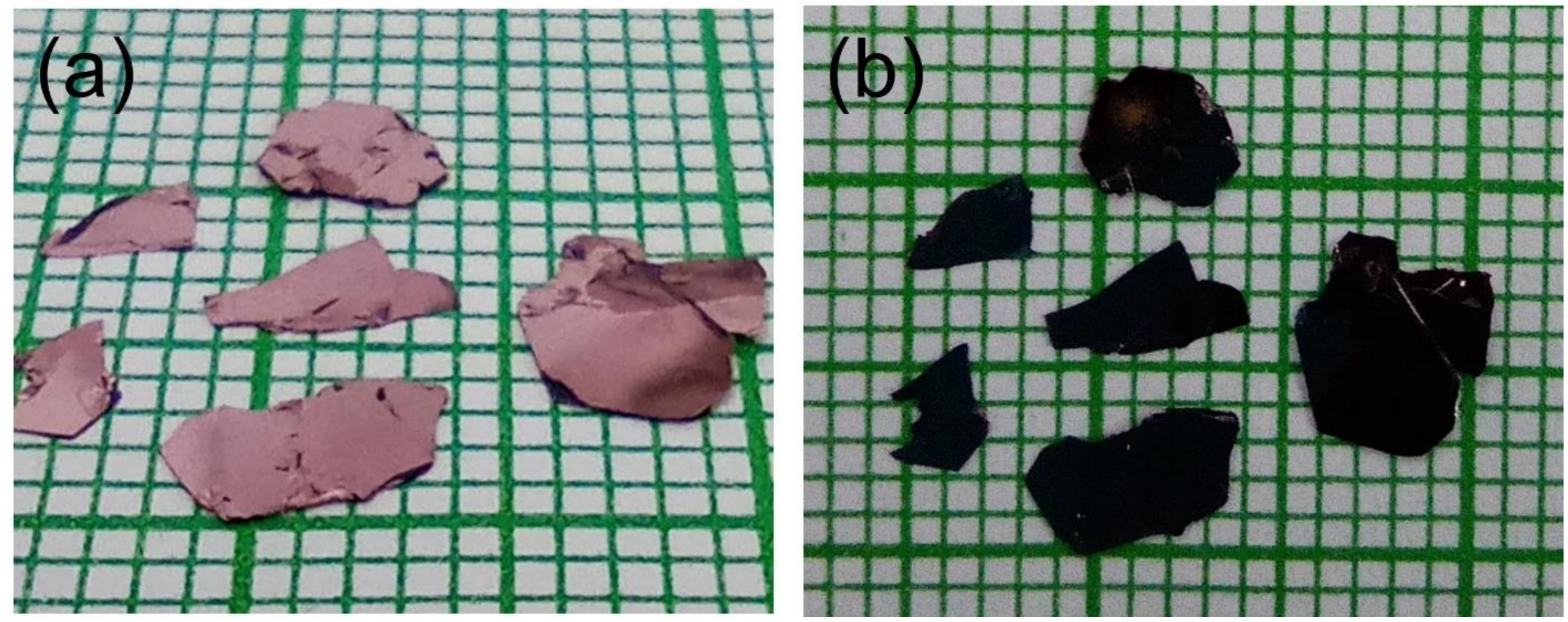

**Figure S1**: (a) and (b) Optical images of as-grown $TiSe_2$ flakes that were used for this study. Large grid has a 1 cm side length.

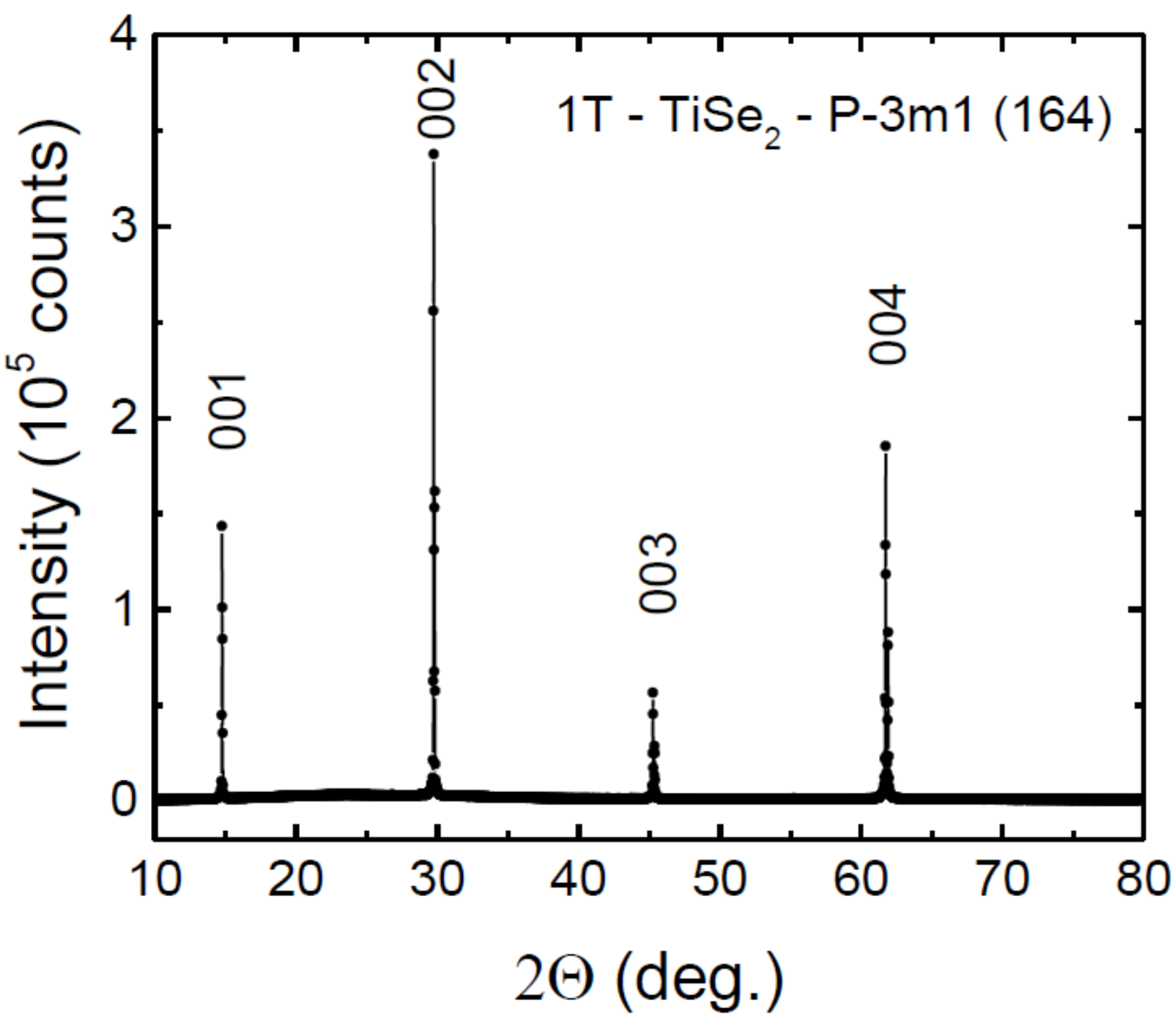

**Figure S2:** X-ray diffraction (XRD) pattern taken in Bragg – Brentano geometry on a single – crystalline platelet of $TiSe_2$. The presence of sharp peaks with indices Miller $(00l)$ implies that the c-axis is perpendicular to the plane of the platelet and that the sample is crystalline. The pattern can be fit to the space group $P\overline{3}m1$.

## 2. Device fabrication

To measure the photocurrent output from a 1T-$TiSe_2$ flake we needed to fabricate electrodes to connect the flake to the external measurement circuit. However, $TiSe_2$ is not very stable in air, (*41*, *42*) and thus exposure to ambient conditions has to be avoided as far as possible. Furthermore, following a standard fabrication route using electron beam lithography is detrimental to the crystal since this leads to the $TiSe_2$ being exposed to the solvents involved in the fabrication process which degrade the crystal and result in the formation of visible large-scale defects. Thus, a new method of collecting current from the flake after photoexcitation was required. Using graphene flakes, it is possible to circumvent top-electrode fabrication and directly place such electrodes on the flake, as shown schematically in Fig. 1(b)-(c) of the main text. First, using electron beam lithography, we patterned electrodes onto a Si/$SiO_2$ substrate. After the electrodes were fabricated on the Si/$SiO_2$ substrate, 1T-$TiSe_2$ crystal was exfoliated using the standard scotch-tape exfoliation method and transferred between the prefabricated electrodes using a microscope and PDMS film. Having transferred the $TiSe_2$ flake, commercially available graphene was quickly exfoliated using the scotch tape method and transferred onto fresh PDMS film. Few-layer thick graphene flakes with a straight edge were chosen under the microscope and placed such that they overlap with both the $TiSe_2$ flake and the metallic electrode, thereby making an electrical connection between the top surface of the $TiSe_2$ flake and the metallic electrode. The distance between the two electrodes on the $TiSe_2$ flake was carefully controlled using the precision stage. To avoid oxidation of the $TiSe_2$ flake through prolonged air exposure once the device was fabricated, we placed a large few-layer flake of hexagonal Boron Nitride (hBN) on top of this $TiSe_2$-graphene stack. The synthesis conditions for the hBN used can be found in ref. (*43*) of the main text. Placing a flake of hBN on $TiSe_2$ is known to protect it from further oxidation and also raise the CDW transition temperature by up to ~ 45K by stabilizing the CDW phase if encapsulated on both sides.[ref (*36*) of the main text] Fig. 1(d) of the main text shows an optical image of a fully fabricated device, showing the various flakes in the stack while Fig. S3 shows a schematic of the overall fabrication process.

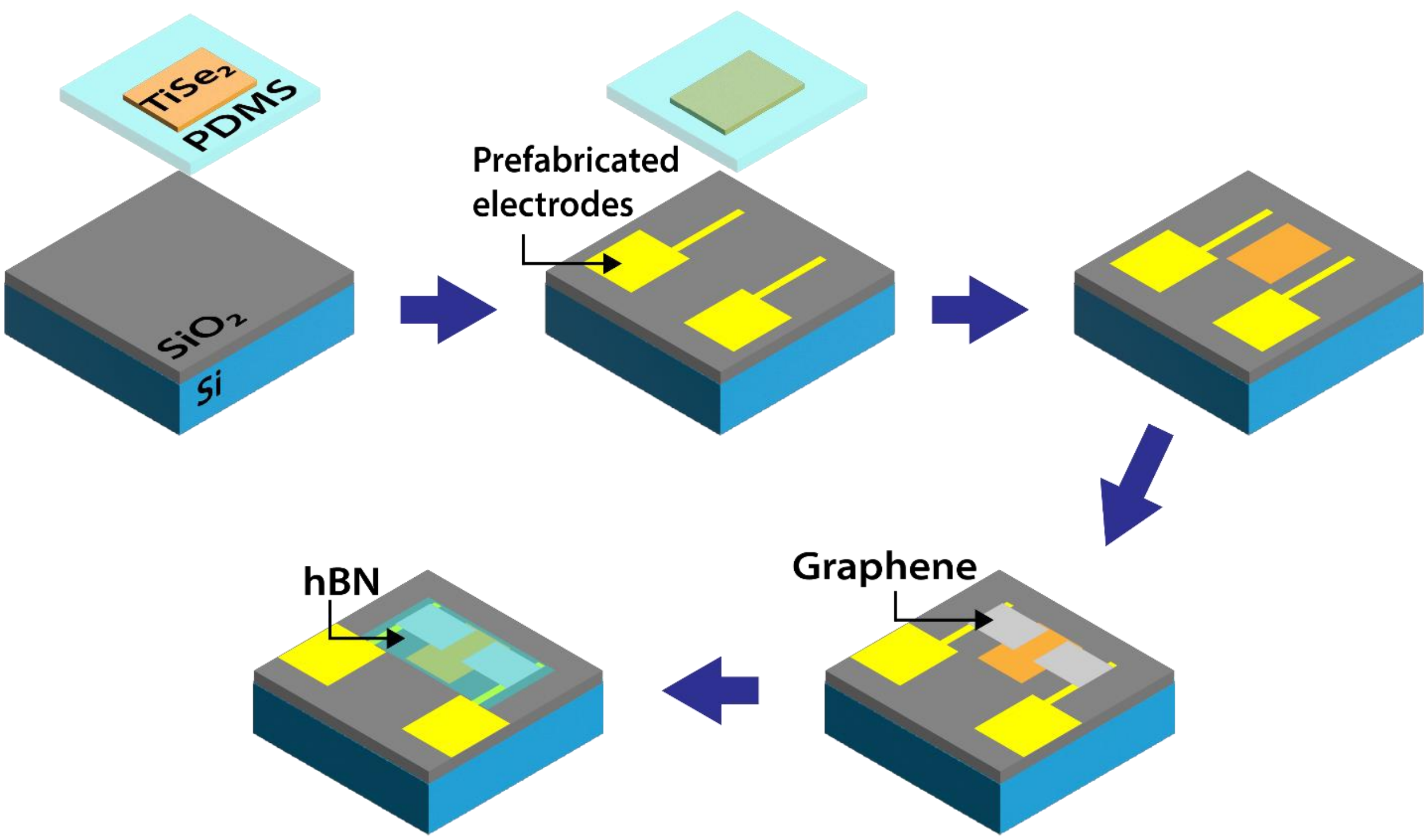

**Figure S3:** Schematic depiction of the PDMS transfer process to make a functional device without top electron beam lithography dependent fabrication. $TiSe_2$ is exfoliated using the scotch tape method, transferred onto a PDMS film, and an appropriate flake is found under an optical microscope. The flake is then transferred onto a Si/$SiO_2$ substrate between two prefabricated gold electrodes. Graphene electrodes are similarly placed using PDMS to connect the $TiSe_2$ flake to the gold electrodes, with an active area of ~10-15$\mu$m between the electrodes. Finally, a few-layer hBN flake is placed on top to encapsulate the device.

The device was then wire-bonded to a chip carrier, placed in the cryostat in the optical setup, and immediately pumped down to a pressure $< 1.0 \times 10^{-6}$ torr. Once the chamber reached the desired pressure, we measured the dark IV characteristics of the device, as shown in Fig. S5. The behavior is linear, indicating Ohmic contacts in the device. This indicates that no significant Schottky barriers are formed at any of the measured contacts, which is one of the potential symmetry breaking processes in such devices leading to extraneous CPGE. The device was then cooled down using liquid nitrogen to 140K to ensure that the system transitions into the CDW state.

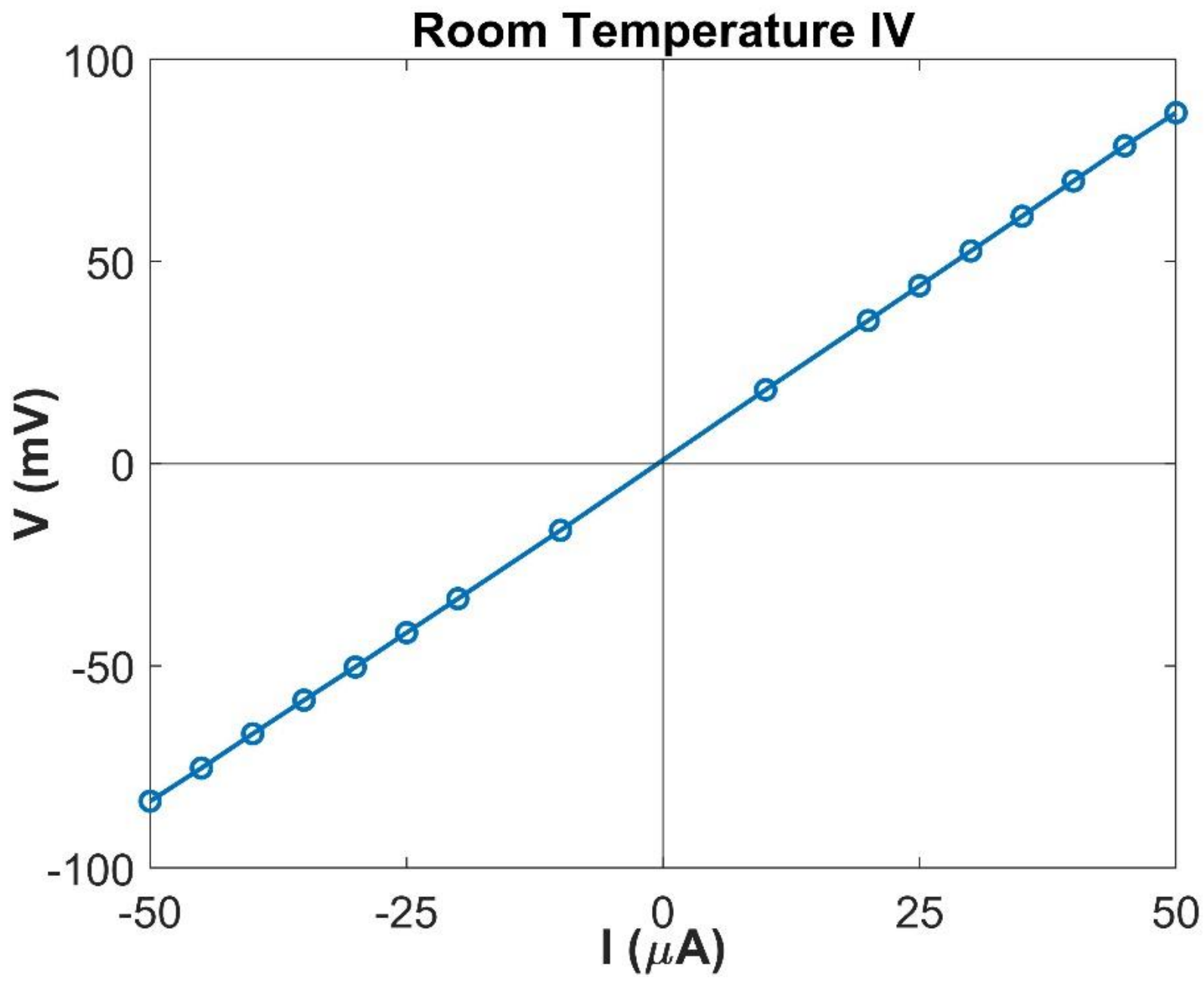

**Figure S5**: Room temperature dark IV curve for a typical device as constructed in Fig.1. The linear IV shows no Schottky barriers.

### 3. Fast CPGE/LPGE measurement setup

In order to get a fast, low-noise measurement of the CPGE or LPGE response of any system, a new measurement technique was developed. Here, the QWP is rotated at some frequency $\omega$ (~160 Hz) using a high-speed motor controlled electronically and the frequency of rotation is input as reference to the lock-in amplifier. Then the current (or voltage) measured at the second harmonic ($2\omega$) gives us the CPGE response of the device, while the fourth harmonic ($4\omega$) response gives us the LPGE which is a consequence of equation (1) of the main text. This is a low noise, fast technique to quickly measure the CPGE (or LPGE) response of devices across various parameter spaces such as wavelength, applied bias, temperature, etc. In this particular case, measuring the PGE using the standard photocurrent measurement is noisy, since the turbopump connected to the cryostat introduces vibrations to the setup, but the usage of this fast-rotation technique helps mitigate this noise and get a low-noise PGE measurement. (See ref. 34 of the main text for more details)

### Supplementary section 1: CPGE and $\gamma$ tensor analysis

Circular photogalvanic effect is a technique highly sensitive to the symmetry of the material under study. To produce such a response, two photons interact with the material system to produce a DC response which is different under left- and right-circularly polarized light. To formally describe CPGE response of a material, consider the material being illuminated by light, the electric field component of which can be described as:

$$E_j(\vec{r},t) = E_j e^{i\omega t + i\vec{q}.\vec{r}} + E_j^* e^{i\omega t - i\vec{q}.\vec{r}} \tag{S1}$$

Then the material response can be written in powers of electric field as a current density given by the expression:

$$j_l(\vec{r},t) = \sigma_{lm} E_m e^{i\omega t + i\vec{q}.\vec{r}} + \sigma_{lmn}^{(2')} E_m E_n e^{2i\omega t + 2i\vec{q}.\vec{r}} + \sigma_{lmn}^{(2)} E_m E_n^* + \cdots \tag{S2}$$

Here the first term contains the linear optical conductivity $\sigma_{lm}$ and thus describes the linear optical response of a material, the second term which oscillates at $2\omega$ or twice the incident photon frequency gives the second harmonic response of the material, while the third term describes the lowest order DC response of the material. This DC term is the relevant term for the photogalvanic response, and can be Taylor expanded with respect to $\vec{q}$ as:

$$j_l^{DC} = \sigma_{lmn} E_m E_n^* + \sigma_{lmnk}(\omega,\vec{q}) q_k E_m E_n^* + \cdots \tag{S3}$$

$$= \lambda_{lmn}(\omega) E_m E_n^* + T_{lmnk}(\omega,\vec{q}) q_k E_m E_n^* + \cdots \tag{S4}$$

Where $\lambda_{lmn}(\omega)$ gives the $q-$independent third order nonlinear conductivity tensor describing CPGE, while $T_{lmnk}(\omega,\vec{q})$ describes a fourth rank tensor required to describe $q-$dependent DC responses such as photon drag effect, quadrupolar CPGE, etc.

Based on the symmetry of the tensor elements of $\lambda_{lmn}(\omega)$, we can decompose the $q-$independent response as:

$$j_l^{DC}(\vec{q}=0) = \frac{1}{2}\lambda_{lmn}^{symmetric}(\omega)[E_m E_n^* + E_m^* E_n] + \frac{1}{2}\lambda_{lmn}^{antisymmetric}(\omega)[E_m E_n^* - E_m^* E_n] \tag{S5}$$

$$= j^{LPGE} + j^{CPGE} \tag{S6}$$

Furthermore,

$$j_l^{CPGE} = \frac{1}{2}\lambda_{lmn}^{antisymmetric}(\omega)[E_m E_n^* - E_n E_m^*] = \gamma_{lk} i \epsilon_{kmn}[E_m E_n^* - E_n E_m^*] \tag{S7}$$

$$= i\gamma_{lk}\left[\vec{E}\times\vec{E}^*\right]_k \tag{S8}$$

The tensor $\gamma_{lk}$ is only non-zero for crystals belonging to the 18 gyrotropic point groups as listed in the main text: $C_1, C_2, C_3, C_4, C_6, C_s, C_{2v}, C_{3v}, C_{4v}, C_{6v}, D_2, D_4, D_{2d}, D_3, D_6, S_4, T$ and $O$.

In the case of 1T-TiSe$_2$, the room temperature space group of the material is $P\bar{3}m1$ while the CDW ground state has the space group $P\bar{3}c1$, both of which have the point group $D_{3d}$. Since $D_{3d}$ is inversion symmetric,

$$\gamma_{D_{3d}} = \begin{pmatrix} 0 & 0 & 0 \\ 0 & 0 & 0 \\ 0 & 0 & 0 \end{pmatrix} \tag{S9}$$

For the proposed optically induced space group of the CDW phase $P321$, the point group symmetry is $D_3$. The gyrotropy tensor $\gamma$ for this point group looks as:

$$\gamma_{D_3} = \begin{pmatrix} \gamma_{xx} & 0 & 0 \\ 0 & \gamma_{xx} & 0 \\ 0 & 0 & \gamma_{zz} \end{pmatrix} \tag{S10}$$

This does not allow coupling of in-plane fields to in-plane currents due to the presence of an in-plane $C_2$ symmetry forbidding off-diagonal tensor elements.

Finally, if we consider that due to the Coulombic screening of the free carriers generated through optical excitation leads to a breakdown of the symmetry along the $c-$axis (or the $z-$axis in the lab frame), the resultant point group reduces from $D_3$ to $C_3$ and the gyrotropy tensor reads as:

$$\gamma_{C_3} = \begin{pmatrix} \gamma_{xx} & -\gamma_{xy} & 0 \\ \gamma_{xy} & \gamma_{xx} & 0 \\ 0 & 0 & \gamma_{zz} \end{pmatrix} \tag{S11}$$

which allows the coupling of in-plane electric fields to in-plane currents due to the presence of the off-diagonal tensor elements, resulting in a non-zero CPGE being observed in the material.

**Supplementary Section 2: Derivation of non-equilibrium CPGE equation**

Consider a pumped three-level system, with the ground state being state 1, state 2 is the lowest excited state, and state 3 is some high energy state. Optically pumping carriers from state 1 to state 3 happens with a probability $P_{13} = P_e$. The carriers then decay from state 3 to state 2 with a lifetime $T_{32}$ while the lifetime of carriers in state 2 is $T_{21}$. To have a nonequilibrium population in state 2 as a result of the optical pumping, we require $T_{21} \gg T_{32}$. Since the carrier population in the system $N$ is conserved, we have the constraint $N = N_1 + N_2 + N_3$. Then, the populations in the three states follow the following kinetic equations:

$$\frac{dN_3}{dt} = P_e(N_1 - N_3) - \frac{N_3}{T_{32}} \tag{S12}$$

$$\frac{dN_2}{dt} = \frac{N_3}{T_{32}} - \frac{N_2}{T_{21}} \tag{S13}$$

$$N = N_1 + N_2 + N_3 \tag{S14}$$

Solving these coupled equations, we get:

$$\frac{N_2 - N_1}{N} = \frac{(T_{21} - T_{32})P_e - 1}{(T_{21} + 2T_{32})P_e + 1} = \frac{(1-a)P_eT_{21} - 1}{(1+2a)P_eT_{21} + 1} \tag{S15}$$

where $a = \frac{T_{32}}{T_{21}} = \frac{N_3}{N_2}$. For a stable nonequilibrium population to be achieved in state 2, we need $a \ll 1$. The equation can be rewritten in terms of a normalized intensity $I = P_e t$ as:

$$\frac{N_2 - N_1}{N} = \frac{(1-a)I_T - 1}{(1+2a)I_T + 1} \quad \text{(S16)}$$

This gives us the first term of the equation (2) of the main text, where the light intensity is leading to a nonequilibrium population in the conduction band of $TiSe_2$, leading to a disruption of the excitonic correlations and therefore the CDW phase of the system.

However, once a steady state population has been created in the conduction band, any more driving will result in generation of CPGE which depends linearly on intensity since $j_{CPGE} = \gamma(\vec{E} \times \vec{E}^*)$. Thus, the two terms lead to a combined effect of the CPGE following the equation (2) of the main text.

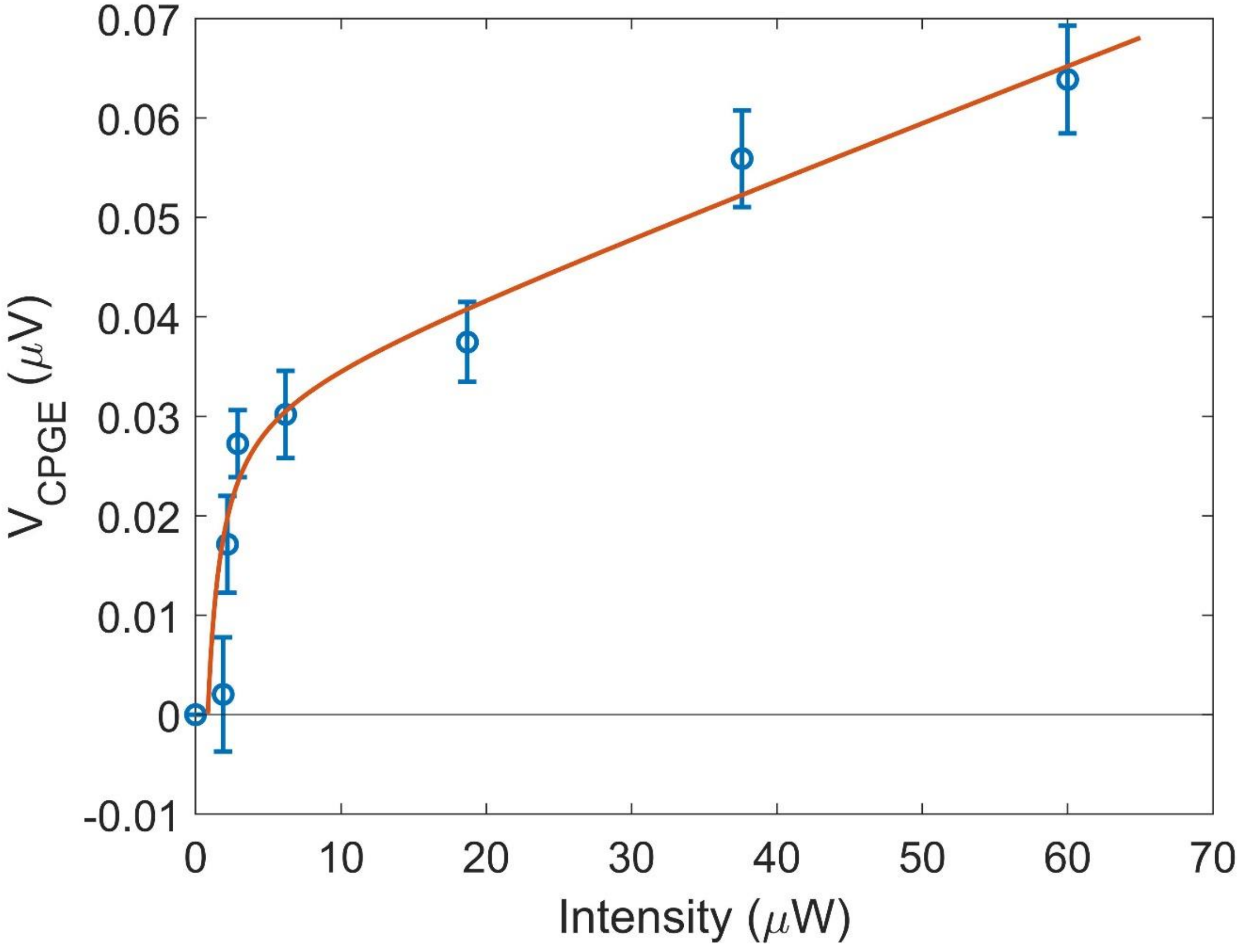

**Figure S6**: Intensity dependent CPGE curve for a second device also showing optically triggered behavior.